\documentclass[traditabstract]{aa}
\usepackage{natbib}
\bibpunct{(}{)}{;}{a}{}{,}
\usepackage{graphicx}
\usepackage{txfonts}
\usepackage{rotating}

\newcommand{\msunyr}{\ensuremath{\mathrm{M}_{\odot}{\rm yr}^{-1}}}   
\newcommand{\kms}{\ensuremath{{\rm km\,s^{-1}}}}                   
\newcommand{\K}{\ensuremath{{\rm K}}}                   
\newcommand{\msun}{\ensuremath{\mathrm{M}_{\odot}}}   
\newcommand{\lsun}{\ensuremath{\mathrm{L}_{\odot}}}                  

\newcommand{\lstar}{\ensuremath{\mathit{L}_{\star}}}                 
\newcommand{\mdot}{\ensuremath{\dot{M}}}                             
\newcommand{\rstar}{\ensuremath{\mathit{R}_{\star}}}                 
\newcommand{\teff}{\ensuremath{\mathit{T}_{\rm eff}}}                
\newcommand{\vinf}{\ensuremath{v_{\infty}}}                          
\newcommand{\vinfi}{\ensuremath{v_{\infty,1}}}                          
\newcommand{\vinff}{\ensuremath{v_{\infty,2}}}                          

\newcommand{\mdotopt}{\ensuremath{\dot{M}_\mathrm{opt}}}                             

\begin{document}

\title{The bi-stability jump as the origin for multiple P-Cygni absorption components in Luminous Blue Variables}
\author{J. H. Groh\inst{1}
\and
J.~S. Vink\inst{2}
}
\institute{
Max-Planck-Institut fuer Radioastronomie, Auf dem Huegel 69, D-53121 Bonn, Germany; \email{jgroh@mpifr.de}
\and Armagh Observatory,  College Hill, Armagh BT61 9DG, Northern Ireland;  \email{jsv@arm.ac.uk}
}

\authorrunning{Groh \& Vink }
\titlerunning{Multiple P-Cygni absorptions in LBVs explained by the bi-stability}

\date{Received  / Accepted }

\abstract{Luminous Blue Variables (LBVs) oftentimes show double-troughed absorption
in their strong H$\alpha$ lines, which are as yet not understood. 
Intriguingly, the feature has also been 
seen in the interacting supernova SN 2005gj, which was for this reason suggested to have 
an LBV progenitor. Our aims are to understand the double-troughed absorption feature
in LBVs and investigate whether this phenomenon is related to wind variability.
To this purpose, we perform time-dependent radiative transfer modeling using {\sc cmfgen}.
We find that {\it abrupt} changes in the wind-terminal velocity -- as expected from 
the bi-stability jump -- are required to explain the double-troughed absorption 
profiles in LBVs. This strengthens scenarios that discuss the link between LBVs and SNe
utilizing the progenitor's wind variability resulting from the bi-stability jump.  
We also discuss why the presence of double-troughed P-Cygni components 
may become an efficient tool to detect extra-galactic LBVs and how to  analyze their mass-loss 
history on the basis of just one single epoch of spectral observations.}


\maketitle

\section{\label{intro}Introduction} 
 
Luminous Blue Variables (LBVs) are unstable massive stars whose evolutionary
state is under heavy debate. While they have been thought to be in a short, transitory phase between an O-type and Wolf-Rayet (WR) star \citep{hd94}, recent observations suggest that LBVs might actually be the final stage before the star explodes as a type-II SN \citep[e.\,g.][]{kv06,smith07,galyam09}.

LBVs are characterized by severe mass loss ($\sim10^{-6}-10^{-3}~\msunyr$), slow winds (terminal speeds $\sim50-500~\kms$), 
photometric and spectroscopic S Doradus-type variations on timescales of years, and are often surrounded by circumstellar nebulae \citep{hd94}.
During the S Dor-type variations, the stellar radius changes by up to a factor of 10, while the effective temperature changes in the range $8\,000-25\,000~\K$ \citep{vg01}. There is evidence for variability in the bolometric luminosity of up to a factor of two in the few LBVs that have been analyzed in detail with modern radiative transfer codes \citep{ghd09,clark09}.

During the S-Dor variations, LBVs may eventually cross the so-called bi-stability limit \citep{vink02,ghd11}, which occurs in line-driven winds at around $\teff\sim21\,000K$ \citep{lamers95,vink02}. Model predictions by \citet{pauldrach90} suggested that wind bi-stability  plays an important role in a shell-ejection scenario of P\,Cygni, in the sense that small variations ($\sim5\%$) in $\lstar$ or $\rstar$ cause significant changes in the wind properties (see also \citealt{najarro97}; \citealt{hillier98}).

Today, bi-stability refers to mass loss and wind-velocity 
variations resulting from changes in the Fe ionization  
\citep{vink99}.  Given that the defining characteristic of LBVs are their S-Dor variations 
\citep{hd94}, there is an expectation that LBVs  could change their mass-loss properties due to the bi-stability whilst they 
travel back and forth across the upper Hertzsprung-Russell diagram 
\citep{vink02,lamers97}. Indeed, \citet{ghd11} confirmed that the prototype LBV AG Carinae crosses the bi-stability limit, but were unable to identify whether a gradual change or a jump in the wind properties occurs.

\begin{figure} 
\resizebox{0.96\hsize}{!}{\includegraphics{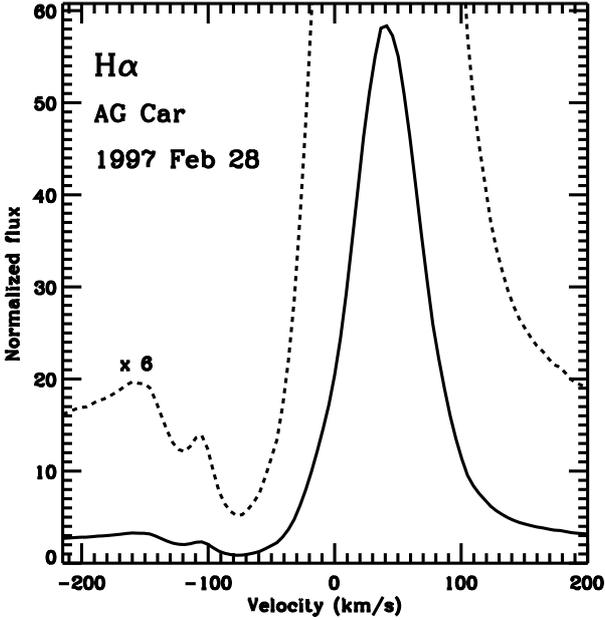}}
\caption{\label{fig1}Observed spectrum of the LBV AG Car around H$\alpha$ obtained in 1997 Feb 28, from \citet{stahl01}. The dashed line shows a magnified view (factor of 6) of the double-troughed P-Cygni absorption. }
\end{figure}

It is also the property of having variable mass loss and wind terminal speed that first revealed LBVs as candidate supernova (SN)
progenitors (Kotak \& Vink 2006). This assertion
was strengthened by the detection of double-troughed 
P\,Cygni absorption components in the narrow H$\alpha$ portion of the interacting supernova SN2005gj 
\citep{trundle08}.
Such multiple-absorption components have been known, for several decades, to exist in H$\alpha$ spectra of confirmed LBVs 
 \citep[Fig.~\ref{fig1}; e.\,g., ][]{stahl83,stahl01,stahl03,leitherer94}. The phenomenon might be the result of  shells ejected by the LBV, or caused by radiative transfer effects \citep{hillier_lbvreview_92}, or due to variable winds, as would be expected from the physics of the bi-stability jump.

The main aim of this Letter is to explain the origin of the multiple P-Cygni absorption components seen in LBVs.  Explaining this phenomenon is particularly relevant as it is as yet far from proven that LBVs really are SN progenitors.
A number of hints in this direction have been suggested, but they all have their specific issues (see e.g. \citealt{dwarkadas11} and references therein). Most significantly, stellar evolution theory 
does not predict the cores of LBV stars to be advanced enough to be anywhere close to 
core-collapse \citep{langer94,meynet03}. It is thus important to scrutinize 
all suggestions involving a direct link between LBVs and SNe. Before being able to attack the line profiles seen in SN2005gj, which involve the complexities inherent to both progenitor mass loss and the subsequent SNe, it seems logical to first attempt to explain the more widespread existence of multiple absorption components in LBVs themselves. 

\section{\label{model}Radiative transfer modeling method} 

We employ a modified version of the spherical symmetric, fully line-blanketed, non-LTE 
radiative transfer code {\sc cmfgen} \citep{hm98} to analyze the effects of 
time-dependent wind parameters on the stellar spectrum.  
Each model is specified by the hydrostatic core radius \rstar, luminosity \lstar, 
mass-loss rate \mdot, wind volume filling factor $f$, wind terminal velocity \vinf, and 
chemical abundances $Z_i$ of the included species. 
We compute time-dependent models with  the following stellar and wind parameters: $\teff$ in the range of $13000$ to $25000~\K$, $\mdot$ in the range of 0.2 to $2\times10^{-5}~\msunyr$, $M=60~\msun$, $\lstar=10^6~\lsun$, $f=0.25$, and 
chemical abundances typical of LBVs, with a He mass fraction of Y=0.62,  N/N$_\odot$=7, C/C$_\odot=0.08$, O/O$_\odot=0.025$, and 
solar abundance for the iron group elements.  These parameters are roughly characteristic for AG\,Car.

Here we briefly describe our implementation of time-dependent density and velocity structures.  The reader is referred to \citet{groh08_clumping} and Groh \& Hillier (2011, in prep) for further details.  At the moment, no time dependence is accounted for in the  solution of the radiative transfer and rate equations, since the recombination timescale is much shorter than the flow timescale for the wind regions studied here.
As the wind hydrodynamics is currently not yet solved for, in order to account for time-dependent 
outflows, we allow for arbitrary $v(r)$ and $\rho(r)$ stratifications as inputs into {\sc cmfgen}.  We consider cases when the star crosses the bi-stability limit, evolving between two epochs: the initial epoch 1, when $\teff=23\,000~K$ and a fast wind is present with $\vinfi$, and epoch 2, when $\teff$ decreases and a slow wind is present, according to $\vinff < \vinfi$. For simplicity, we keep $\mdot$ fixed, so that $\rho(r)$ is computed following the equation of mass conservation.

We explore two  scenarios as to whether the double P-Cygni profiles can be caused by the 
time variability of wind parameters: a gradual change in $\vinf$, or an abrupt instantaneous 
variability. 

\begin{figure}
\resizebox{1.00\hsize}{!}{\includegraphics{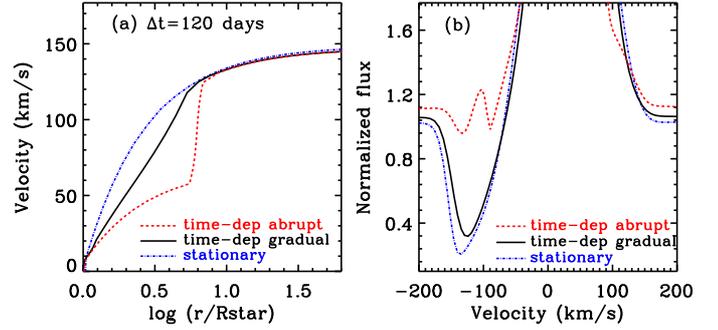}}
\caption{\label{fig2}{\it (a)}: wind velocity law for a time-dependent model at  $\Delta t=120$~days with a gradual change in $\vinf$(t) (black solid line) and with a abrupt change in $\vinf$(t) (red dashed line), in comparison to a steady-state model (blue dash dotted line). 
{\it (b)}: corresponding H$\alpha$ line profiles. }
\end{figure}

For models with gradual changes in $\vinf (t)$, for simplicity we assume that, starting at $t=0$, $\vinf$ decreases linearly with time. For 
a given time $\Delta t$ after $t=0$, we compute the distance traveled by material ejected between $t=0$ and $\Delta t$. The result is a velocity law that increases linearly with the distance up to a point where the initial wind is reached. From that point onwards, $v(r)$ follows the standard $\beta$-type, asymptotically reaching $\vinfi$ (Fig.~\ref{fig2}a). 

For models with an abrupt change, we assume that at $t=0$, $\vinf$ switches instantaneously 
from $\vinfi$ to $\vinff$. To compute a model for a given $\Delta t$, we evaluate the 
distance traveled by the last particles of the initial wind ($r_1$), 
as well as the distance traveled by the first particles of the final wind ($r_2$). 
Following a suggestion by S.P. Owocki (2010, priv. comm.), we assume that the initial wind characterizes 
the region from $r_1$ up to the outer boundary of the computational domain, whilst a $\beta$-type law characterized 
by $\vinfi$ and $\beta_1$ is assumed. 
From the star up to $r_2$, we assume that $v(r)$ is given by the velocity law of the final wind according 
to $\vinff$ and $\beta_2$. 
We further assume that the region in between the two winds, i.e. between $r_2$ and $r_1$, $v(r)$ and $\rho(r)$ 
are linear, and join the two winds smoothly. 
We defer to future hydrodynamical computations for a more accurate treatment of this region, but we 
note that initial tests using somewhat different assumptions may lead to small changes in the resulting P-Cygni 
line profiles. These changes are not anticipated to affect the conclusions reached here.
 
For both gradual and abrupt models, we assume that the stellar wind is at $t=0$ in 
steady-state with $\vinfi=150~\kms$ and $\beta_1=1$ and, as $t\to \infty$, evolves to a final state 
with $\vinff=70~\kms$ and $\beta_2=1$. 
The choice of $\vinf$ values is based on LBV spectral data, whilst the factor of two change in $\vinf$
is adopted from theoretical predictions \citep{vink99}. 
 
\section{\label{abrupt}Time-dependent P-Cygni line profiles: gradual $vs.$ abrupt changes in the wind parameters} 

Figure \ref{fig2}b presents the P-Cygni line profiles calculated for both a gradual and an abrupt variation 
of the wind parameters as a function of time, here computed at $\Delta t=120$~days. 
A gradual change in $\vinf(t)$ causes a linear variation of the velocity with 
distance (Fig.~\ref{fig2}a). At $\Delta t=120$~days, the P-Cygni profile deviates only slightly from a steady-state model with 
the same stellar parameters.  To test the robustness of our results, we computed models for different $\Delta t$, as well as stellar and wind 
parameters all with gradual changes, but we still could not obtain double P-Cygni absorption components.

On the other hand, we found that an abrupt change in $\vinf$ causes significant effects in the wind velocity 
law (Fig.~\ref{fig2}a). While the velocity law of the outer wind is similar to that of the gradually changing as well as steady-state models, the 
inner part of the wind follows a $\beta$-type law asymptotically approaching $\vinff$. 
The abrupt change in $\vinf$ modifies not only the velocity and density structures, but also causes the 
appearance of a double-troughed P-Cygni absorption component (Fig.~\ref{fig2}b). 
The double P-Cygni absorption component arises as a result of the presence of a small amount of material 
with velocities in between $\vinfi$ and $\vinff$, causing a very low H$\alpha$ line opacity between 
these two velocities. Models computed with a variable $\mdot$ across the bi-stability jump show a similar qualitative behavior as the ones shown here, implying the abrupt change in $\vinf$ is the key phenomenon responsible for the double-troughed P-Cygni absorption components. We are not able to tell whether there is a jump in $\mdot$ as well.

\begin{figure}
\resizebox{1.00\hsize}{!}{\includegraphics{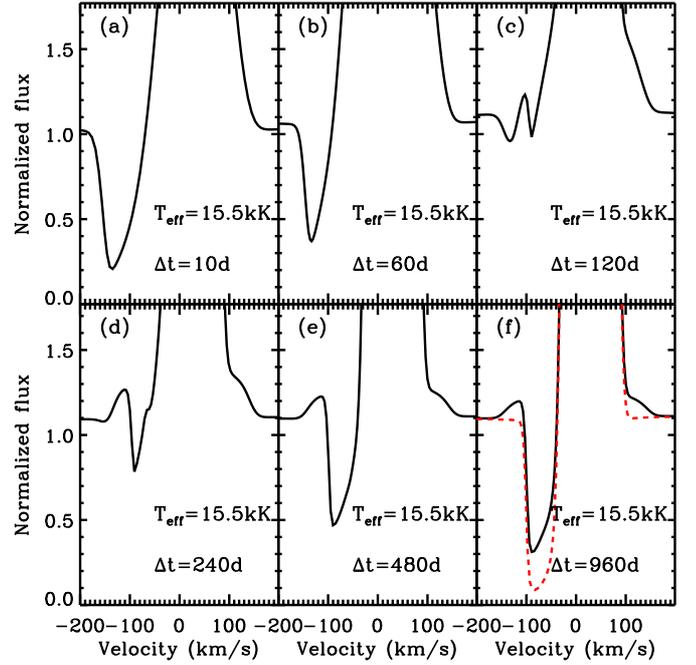}}
\caption{\label{fig3b}Variations of the synthetic H$\alpha$ line profiles as a function of $\Delta t$ for a fixed $\teff=15.5$ kK and $\mdot=9 \times10^{-6}~\msunyr$. The red dashed line in panel {\it (f)} corresponds to the steady-state profile obtained as $t\to \infty$.  }
\end{figure}

Figure~\ref{fig3b} presents the temporal evolution of the H$\alpha$ absorption profile 
for selected $\Delta t$, showcasing how the absorption evolves from the initial fast wind 
in steady-state (panel $a$) to the final slow wind in steady-state (panel $f$, red dashed line). 
We find that the P-Cygni absorption component from the fast wind becomes weaker, as the slow wind takes over, 
disappearing at around $\Delta t=240$~days. 
The P-Cygni absorption component from the slow wind first appears rather weakly at 
around $\Delta t=90$~days, and increases in strength as a function of time.
Note that even when no absorption from the fast wind is seen, its weak emission still 
influences the spectral morphology, even at $\Delta t=960$~days.

So when can the double P-Cygni absorption components be seen after the abrupt change in $\vinf$? 
The duration of the double P-Cygni absorption period depends on the characteristic flow time 
(dependent on $\rstar/\vinf$ and $\beta$) of both winds, as well as on the location of the line 
forming region of interest. 
For the parameter space explored here, H$\alpha$ forms at a distance from $\sim$2 to 10~$\rstar$, and 
the double P-Cygni absorption component is seen from $\Delta t$ around 90 to 180 days, with its maximum 
strength around 120~days. Higher values of $\beta$ increase the flow timescale and  
allow the double P-Cygni absorption components to be observed during a more extended period of time. 

\section{\label{parameter}The dependence of the double P-Cygni absorption profiles on $\teff$ and $\mdot$} 

An abrupt change in $\vinf$ is required to reproduce 
double P-Cygni absorption profiles (Fig. \ref{fig2}). The next relevant question is: for which range of 
stellar and wind parameters does a double P-Cygni absorption profile appear in H$\alpha$?

\begin{figure}
\resizebox{1.00\hsize}{!}{\includegraphics{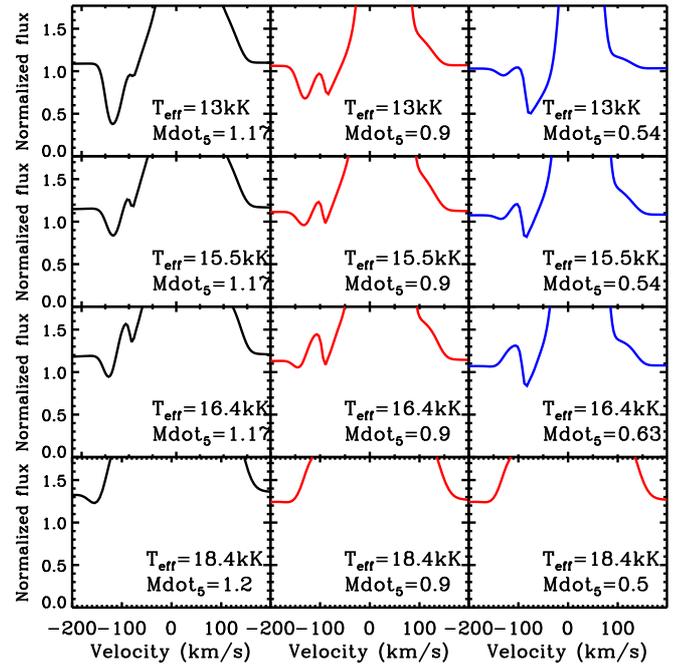}}
\caption{\label{fig3}Variations of the synthetic H$\alpha$ line profiles as a function of $\teff$ (in kK) and $\mdot$ (in units of $10^{-5}~\msunyr$), at  $\Delta t=120$~days. }
\end{figure}

Here we focus on changes in $\teff$ and $\mdot$, keeping the other stellar and wind parameters constant as well as $\Delta t$=120~days. 
We found that the morphology of the double P-Cygni absorption profiles is strongly sensitive 
to both $\teff$ and $\mdot$, and that a zoo of absorption line profiles are predicted by our models (Fig.~\ref{fig3}).

We note that an increase in $\teff$ from $13000~\K$ to $18400~\K$ causes less absorption by the 
initial, fast wind, resulting in weaker high-velocity absorption components. 
Models with $\teff > 18400~\K$ have H ionized throughout the wind, which dramatically reduces 
the population of the n=2 level related to the H$\alpha$ absorption.  
As a consequence, these hotter models show extremely weak or no P-Cygni absorption,
since the time-dependent velocity structure is not probed by H$\alpha$ if H is ionized throughout the wind.

In addition, our model predict weak blue-shifted emission between the 
two absorption components. This blue-shifted emission arises in the initial, fast 
wind, from non-core rays. According to our models, the intensity of the blue-shifted emission increases as $\teff$ increases. 

\begin{figure}
\resizebox{0.90\hsize}{!}{\includegraphics{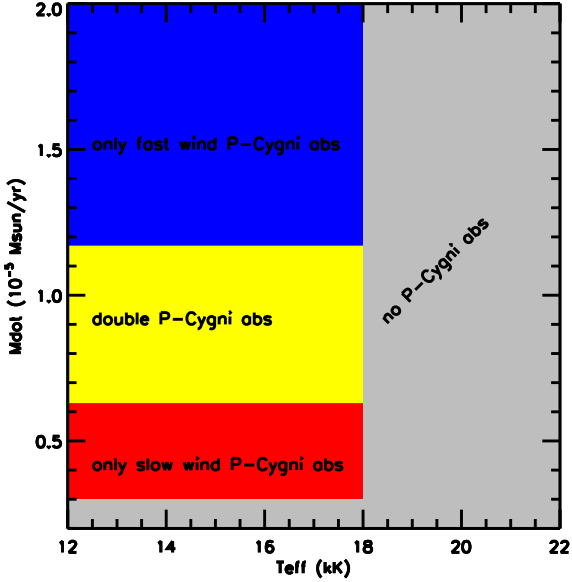}}
\caption{\label{fig4} Morphology of the H$\alpha$ P-Cygni absorption 
component in LBVs as a function of $\mdot$ and $\teff$, at  $\Delta t=120$~days.}
\end{figure}

Perhaps surprisingly, our models predict that, for a given $\lstar$ and $\teff$, double-troughed P-Cygni absorption
appears in H$\alpha$ only for a relatively narrow range of $\mdot$.  We found that there is an optimal value of  $\mdot$ ($\mdot_\mathrm{opt}$) when the contrast between the two absorption components is maximal. The double-troughed P-Cygni absorptions  are present  when $\mdot$ is about $\sim \pm30\%$ around $\mdot_\mathrm{opt}$. For $\teff < 18000~\K$, we found four possible regimes of the H$\alpha$ absorption morphology  (Fig.~\ref{fig4}), depending on $\mdot$:\\
$\bullet~\mdot > 1.3 \mdotopt$:  absorption from the fast wind only, since the emission from the fast wind from non-core rays fills out the absorption from the slow wind; \\
$\bullet~0.7 \mdotopt < \mdot < 1.3 \mdotopt$: double P-Cygni absorption, i.e. absorption from both winds;\\
$\bullet~0.3 \mdotopt < \mdot < 0.7 \mdotopt$: absorption from the slow wind only since, for lower $\mdot$, the fast wind is not optically-thick enough in H$\alpha$ to produce detectable absorption. \\
$\bullet~ \mdot < 0.3 \mdotopt$: no P-Cygni absorption, since both winds are not optically-thick enough in H$\alpha$. 

We note that the ranges given above are only indicative and valid for $\Delta t$=120~days, since the relative location of the four different regions described above varies with $\Delta t$. In reality, the actual observational detection of the double P-Cygni absorption should also depend 
on the spectral resolution and the signal-to-noise ratio of the observations, amongst other factors.

\section{\label{discussion} Discussion}

We demonstrated via time-dependent, radiative transfer modeling with {\sc cmfgen} that abrupt wind-velocity changes are 
needed to reproduce the observed double P-Cygni absorption components seen in a plethora of LBV spectra.  The most promising manner causing such abrupt changes in $\vinf$ is by means of the bi-stability jump, as discussed in \citet{vink02}.

We reinforce that the modeling performed here is not exhaustive, given the vast parameter space to be explored and the model limitations. For instance, the full hydrodynamics and time-dependent solution of the radiative transfer and statistical equilibrium equations are not yet accounted for. Furthermore, we consider an idealized scenario of an instantaneous abrupt change in \vinf, whereas in reality, the changes could occur over extended time periods of days or weeks. Constraints on the timescale could be obtained via time-dependent  
modeling of observational data of LBVs that actually cross the bi-stability limit, e.g. AG Car. These topics will be the subject of future studies. 

We note that the empirical identification of a luminous massive star as a bona-fide member of the 
LBV class is a challenging exercise. This is mainly due to the long timescales 
(years to decades) necessary to monitor the irregular spectral variability. As the detection of double P-Cygni absorption components is
a {\it strong} indication that the candidate object is an LBV undergoing S-Dor type variability 
cycles -- with its abrupt changes in $\vinf$ on timescales of years -- we 
suggest that the sheer presence of such double P-Cygni absorption components may 
become an efficient, novel way to detect Galactic and extra-galactic LBVs. 

We found that the morphology of the double P-Cygni absorption components depends strongly 
on $\mdot$ and $\teff$. This opens up the possibility to constrain these quantities without the lengthy 
process of non-LTE radiative transfer modeling: once $\lstar$ is known, for given $\Delta t$, $\vinfi$, and 
$\vinff$ could be derived empirically. 

If observations at multiple epochs are available, we should also be able to trace the LBV mass-loss history 
using a similar modeling strategy as employed here, as the H$\alpha$ line forms over an 
extended stellar-wind region in LBVs. This will allow us to use the double P-Cygni absorption
components as a tool to {\it look back in time}, and to be able to estimate LBV stellar parameters from 
one {\it single} epoch. Although in this Letter we focused  on H$\alpha$, motivated by the  observational data available,  we note that our time-dependent models predict similar double-troughed P-Cygni absorption in many other spectral lines formed at different wind regions, such as most of the Balmer lines, \ion{Fe}{II}, \ion{N}{II}, \ion{He}{I} $\lambda$10830, among others. Analyses of these lines should lead to a much improved understanding of the mass-loss history of LBVs.

Interestingly, the morphology of the absorption-line profiles seen 
in Fig.~\ref{fig3} closely resemble those found in the interacting SN 2005gj \citep{trundle08}. 
This strengthens the link between LBVs and SNe that based their assertions on the bi-stability mechanism during S\,Dor cycles 
\citep{kv06,trundle08}, and the LBV/SN link in general \citep{smith07,galyam09}. We expect that our results will encourage further observational and modeling efforts to establish under which conditions  LBVs explode as SNe.

\begin{acknowledgements}
We thank John Hillier for making {\sc CMFGEN} available. We acknowledge discussions with Rubina Kotak, Stan Owocki, and Tom Madura. 
JHG is thankful for the warm hospitality during his visit to Armagh Observatory.
\end{acknowledgements}

\end{document}